\documentstyle[11pt,aaspp4]{article}

\newcommand{\HDF}{HDF~J123652+621227}
\newcommand{\HDFx}{HDF~J123656+621221}

\newcommand{\grad}{\mbox{\boldmath $\nabla$}}           %bold face nabla
\newcommand{\simg}{\raisebox{-0.7ex}{\mbox{$\stackrel{\textstyle >}{\sim}$}}}
\newcommand{\siml}{\raisebox{-0.7ex}{\mbox{$\stackrel{\textstyle <}{\sim}$}}}

\begin{document}

\title{
A candidate gravitational lens in the Hubble Deep Field
}

\author{
	David W. Hogg\altaffilmark{1},
	Roger Blandford,
	Tomislav Kundi\'c}
\affil{ \sl
Theoretical Astrophysics, California Institute of Technology, Mail
Code 130-33, Pasadena, CA 91125}
\altaffiltext{1}{\tt hogg@tapir.caltech.edu}

\author{
	C. D. Fassnacht}
\affil{ \sl
Department of Astronomy, California Institute of Technology, Mail
Code 105-24, Pasadena, CA 91125}

\author{
	Sangeeta Malhotra}
\affil{ \sl
Infrared Processing and Analysis Center, California Institute of
Technology, Mail Code 100-22, Pasadena, CA 91125; and NRC Fellow, NASA
Jet Propulsion Laboratory.}

\begin{abstract}
The discovery of HDF~J123652+621227, a candidate gravitational lens in
the HDF, is reported.  This lens may be multiply imaging several
optical sources at different redshifts.  If follow-up spectroscopy of
the lens and the brightest image confirms this hypothesis,
observations of this system alone can be used to obtain an estimate of
the redshift distribution at extremely faint flux levels.
\end{abstract}


\section{Introduction}

Cosmologically distant galaxies ought to act as multiply imaging
gravitational lenses for a fraction $\sim0.002-0.005$ of background
sources (Turner, Ostriker \& Gott 1984; Blandford \& Narayan 1992;
Schneider, Ehlers \& Falco, 1992). This prediction is being borne out
by surveys of flat spectrum radio sources (Patnaik et al 1992; Myers
et al 1995) and optical surveys (Maoz et al 1992; Glazebrook et al
1994; Ratnatunga et al 1995).  The incidence and character of strong
gravitational lenses provide an important constraint on the source
redshift distribution and world model at magnitudes too faint for
direct spectroscopy (Kneib et al 1994; Kochanek 1992).

Recent Hubble Space Telescope (HST) observations of the Hubble Deep
Field (HDF; Williams et al 1995) permit the optical lensing rate to be
estimated in a uniform manner, using a single observation.  The HDF
images are the deepest images in the visible ever taken: U, B, V and
I-band images with point-source detection limits near 27, 29.5, 29.5
and 28.5~mag, respectively.  Approximately 2500 faint ``galaxies'' can
be identified over 4~square arcmin, so the total number on the whole
sky amounts to $\sim9\times10^{10}$~sources, a number roughly thirty
times the product of the local bright galaxy density and the volume of
an Einstein-de~Sitter Universe out to $z\sim3$.  Possible explanations
of this excess include fading (Babul
\& Rees 1992) or merging (Guiderdone \& Rocca-Volmerange, 1991) of
galaxies, counting multiple sub-galactic star formation sites within a
common potential well as individual galaxies (Katz 1992; Colley et al
1996), or extreme cosmological models with large amounts of comoving
volume per unit luminosity distance (Fukugita et al 1990).  An
important means of distinguishing between explanations is to determine
the redshift distribution of these sources.

In the HDF, $\sim3-10$ cases of multiple imaging are expected (e.g.,
Turner et al 1984; Miralda-Escud\'e \& Lehar 1992); the actual number
constrains the redshift distribution of the very faint sources
observed in the HDF relative to brighter populations (such as quasars)
for which both the lensing rate and the redshift distribution are
better known.  This measurement can be used to constrain the redshift
distribution of sources at much fainter levels than the limits of
current spectroscopic surveys.  We have begun a partially automated
search for multiply imaged sources derived from that used in the CLASS
survey (Myers et al 1995).  After inspection of two dozen candidate
lens systems, the most probable case was found to be
\HDF. This system consists of 16 components, all within 5~arcsec of a red,
$F814W=23.3$~mag elliptical galaxy (component~0, hereafter c.~0) at
RA\,$12^h\,36^m\,52^s\!\!.01$, Dec\,$+62^{\circ}\,12'\,27''\!\!.3$
(J2000), identified as the lens candidate (Figure~1). The most
striking of these companion images are a thin, $F606W=26.1$~mag
tangential arc (c.~3) on one side and a $F606W=27.4$~mag counterimage
(c.~1) on the other.  This configuration is seen in other
gravitational lenses, e.g., FSC10214+4724 (Eisenhardt et al 1996).  It
occurs when a source is located close to a cusp of the lens mapping in
the source plane (Blandford \& Narayan 1992; Schneider et al 1992).
In addition to the arc and counterimage, there are a number of other
faint sources surrounding c.~0 which may be multiply imaged.  If this
hypothesis is confirmed by follow-up observations, this lens system
alone will provide significant constraints on faint source redshift
distributions, because the redshifts of all the lensed and unlensed
components nearby c.~0 can be estimated or at least constrained with
lens models.

\section{Models}

A simple lens model is made for the locations and relative
magnifications of cc.~1 and 3, based on a circular disk background
source (denoted A) of radius $r_A$ being lensed by c.~0, with the lens
modeled with a three-parameter, singular isothermal elliptical
potential (Kochanek 1991; Blandford \& Narayan 1992; Schneider et al 1992)
\begin{equation}
\psi(r,\theta) = b\,r + \eta\,r\,\cos 2(\theta-\theta_{\eta})
\end{equation}
Here $(r,\theta)$ is a coordinate system on the sky with the lens
center at $r=0$; $\grad\psi$ is the deflection angle; $b$ is the
asymptotic critical radius, or radius of the Einstein ring in the
absence of any ellipticity or external shear; $\eta$ sets the
ellipticity of the potential and $\theta_{\eta}$ sets the orientation.
The best-fit model, which reproduces the observed positions, relative
fluxes, and relative tangential lengths of cc.~1 and 3 well, is shown
in Figure~2.  In addition to the quality of the fit, this model is
also compelling because the derived orientation of the surface
potential agrees within measurement error ($\sim10$~deg) with the
position angle of c.~0 and because c.~2 can also be accommodated in
this model by introducing a second source, A$'$, close to A but with a
different color.  Its counterimage is near c.~1 but too faint to be
observed.  In the circular-source model, c.~1 is tangentially
elongated while it is marginally extended in the radial direction in
the HDF images.  This discrepancy can be ameliorated by making the
background source elliptical.  In addition, cc.~1 and 3 have slightly
different colors.  We attribute this to the differential magnification
of spatially distinct (and hence differently colored) regions of the
source given that it lies very close to a caustic in the source plane.
Such differential magnification is important in arc-counterimage
systems (cf., Eisenhardt et al 1996).  Both of these problems would be
eliminated if c.~3 were modeled as a ``naked cusp'' (Schneider et al
1992) in which case there would be no counterimage and c.~1 would be a
foreground galaxy, probably associated with c.~0, and possibly
providing some of the large shear required to make the naked cusp
geometry.

Parameter $b$ is related to 1-D velocity dispersion $\sigma_v$ by
$(\sigma_v/c)^2=b/(4\pi\beta_A)$, where $\beta_A$ is the ratio of the
angular diameter distance from lens to source to the angular diameter
distance from observer to source, which depends on cosmological
parameters and lens and source redshifts.  In this model
$\sigma_v\geq240~{\rm km\,s^{-1}}$, with $\sigma_v$ tending to its
lower bound as $\beta_A\rightarrow 1$, i.e., as the redshift $z_0$ of
c.~0 becomes much smaller than the redshift $z_A$ of source A.

Looking out to larger angular radii from c.~0, it is striking that the
two reddest objects (cc.~9 and 11) in the vicinity of c.~0 have the
same colors as one another, lie at comparable radii from c.~0 but on
opposite sides, along the major axis of c.~0, the line of maximum
lensing probability (Schneider et al 1992).  It is possible to
generalize the lens model to accomodate this pair and at least one
other by adding additional sources at different redshifts and angular
positions behind c.~0.  The lens potential is augmented with a two
parameter, external shear (Kochanek 1991) to make
\begin{equation}
\psi(r,\theta) = b\,r + \eta\,r\,\cos 2(\theta-\theta_{\eta})
 + \gamma\,r^2\,\cos 2(\theta-\theta_{\gamma})
\end{equation}
which allows for a potential ellipticity which
varies with radius.  For each new source $i$ an additional,
independent parameter is the angular diameter distance ratio
$\beta_i$.  In decreasing order of probability of being multiply
imaged, several pairs of components are discussed below:
\begin{enumerate}
\item
cc.~9 and 11 both have colors of $F606W-F814W=0.9$~mag, placing them
among the reddest 300 images in the HDF.  Their source is designated
B.  The images can be fit in position and relative magnification as a
gravitational lens pair with only minimal changes to the base lens
potential with $\beta_B=2.31\beta_A$.
\item
cc.~5 and 7, assigned source C, can be fit in position but less well
(to within a factor $\sim 2$) in relative magnification with
$\beta_C=1.58\beta_A$.  The addition of new potential parameters allow
this observable to be reproduced accurately (albeit with a decrease in
the simplicity of the model).
\item
We find that the positions of cc.~10 and 12 cannot be fit to better
than $0.5$~arcsec in any simple model.  We therefore propose that they
do not comprise a gravitational lens pair but rather have separate
single sources with $\beta<1.2\,\beta_A$.  This is consistent with these
components having significantly different $F450W-F606W$ colors.
\item
The positions of cc.~8 and 13 can be reproduced within the model using
a single source.  However, the magnification ratio differs from the
observed value by a factor $\sim25$.  We therefore conclude that
either these are two independent sources or that c.~8 is blended with
a brighter foreground source.  Either explanation is consistent with
the observed color differences, and in either case the majority of the
flux from c.~8 must be coming from $\beta<0.9\,\beta_A$.
\item
In the absence of obvious counterimages of similar color, we expect
that cc.~4, 6, 14 and 15 are not multiply imaged.
\end{enumerate}
Figure~3 shows a model fit to all the likely multiple
image pairs from the above list.  If indeed cc.~9 and 11 are a pair
multiply imaged by c.~0, then the velocity dispersion of c.~0 rises to
$\sigma_v\geq 380~{\rm km\,s^{-1}}$ with $\sigma_v$ tending to its
lower limit as $\beta_B\rightarrow 1$.  However, because cc.~9 and 11
are detected in the U band, which ought to lie longward of the Lyman
continuum, source B may have $z_B<3$ (Guhathakurta, Tyson \& Majewski
1990; Steidel et al 1996).

The colors of the elliptical lens galaxy c.~0 can be compared with
spectral energy distributions of local elliptical galaxies (Coleman,
Wu \& Weedman 1980) to estimate a photometric redshift.  For this
purpose, visible fluxes from the HST data and a preliminary
near-infrared flux of $K\approx 19.9$~mag (Soifer, B. T., Matthews,
K. \& Armus, L., private communication; Cowie, L. L., private
communication) were used; the best-fit redshift for c.~0 is
$z_0=1.0$--$1.3$. (The uncertainty is somewhat greater because we have
ignored evolution.)  We will adopt $z_0\sim 1$, $H_0=100\,h~{\rm
km\,s^{-1}\,Mpc^{-1}}$ and $(\Omega,\Lambda)=(0.05,0.0)$ in what
follows.  When $z_0\sim 1$, $\beta\,\siml\, 0.5$ for any source
redshift, implying $\sigma_v\geq 340~{\rm km\,s^{-1}}$ if only cc.~1
and 3 are a gravitational lens pair and $\sigma_v\geq 540~{\rm
km\,s^{-1}}$ if cc.~9 and 11 are also a lens pair.  Using the
Faber-Jackson (Faber \& Jackson 1976) and fundamental plane (Bender,
Burstein \& Faber 1992) relations we estimate the velocity dispersion
to be $\sigma_v=250 {\rm km\,s^{-1}}$ if the elliptical galaxy is at
z=1 and $\sigma_v=325 {\rm km\,s^{-1}}$ if the galaxy is at
z=1.5. These numbers are dependent on cosmography and k-corrections
which do not take into account evolution.  Thus, the properties of
local elliptical galaxies suggest that c.~0 is not massive enough to
account for these wide-separation multiple image systems (cc.~9, 11
and 5, 7).  On the other hand, this object is selected for multiple
imaging, not flux, so it may be associated with a much larger mass
distribution than its central light profile would suggest.  It is
possible that c.~0 is surrounded by a poor cluster like Q~0957+561
(Angonin-Willaime, Soucail \& Vanderreiest, 1994) but that the other
members galaxies are either too faint or too spread out in solid angle
to be identified with c.~0.

\section{Predictions}

The low-lens-mass hypothesis (i.e., cc.~1 and 3 comprise the only
gravitational lens pair) requires $z_A-z_0\,\simg\, 0.5$ while the
high-lens-mass hypothesis (cc.~9, 11, and 5, 7 are also pairs)
requires $\beta_B\sim 2.3\,\beta_A$, only possible if
$z_A-z_0\,\siml\, 0.5$.  If the observed redshifts of cc.~0 and 3,
both of which may be obtained in spectroscopic observations currently
underway with the Keck Telescope (Cohen, J. G., private communication;
Cowie, L. L., private communication; Koo, D. C., private
communication; Steidel, C. C., private communication) are consistent
with high lens mass, there will be a small but fair sample of distant,
faint galaxies (sources A, B and C) whose redshifts and luminosities
can be inferred and with which constraints on faint source redshift
distributions can be derived by, e.g., maximum likelihood methods.  It
is emphasized that sources close to c.~0 which are not multiply imaged
are just as important in constructing such constraints as those that
are multiply imaged.  Furthermore, we note that in the high-lens-mass
case, two of the multiply imaged sources (B and C) would be too faint
to be detected were they not magnified, so some information is
obtained about extremely faint sources, for which we have no {\em a
priori\/} distance information.  Although the distance ratios
$\beta_i$ are relatively insensitive to the cosmography (Schneider et
al 1992), it is conceivable that a multi-source lens might be found
where redshifts are measured for several images at different radii or
where the very existence of images at large radii allows us to
constrain or even determine the world model (Soucail \& Fort 1991).
Finally, not only does the gravitational lens magnify the flux of the
background sources, it also elongates them tangentially; the derived
radii of sources A, B and C under the trial model are $190$, $<8$ and
$<25\,h^{-1}$~pc respectively.  This may provide a strong clue as to
the nature of these objects.

% for beta=0.43, 0.01 arcsec is 62.5 h^{-1} pc

There are three possible further tests of the lensing hypotheses.
First, multiply imaged components must have similar infrared colors
and optical--radio spectral indices.  Observations are underway to
obtain ultra-deep imaging of the HDF field at 2$\mu{\rm m}$
(Neugebauer, G., et al., in preparation) and 5~GHz (Fomalont, E. B.,
private communication).  Greater sensitivity in the infrared should
ultimately be achieved with Keck adaptive optics or HST/NICMOS.
Secondly, it is surprising that in this field, \HDF\ alone shows
evidence for multiply-imaging many sources.  We have suggested that
c.~0 may be either a more massive galaxy than is apparent from the HDF
image, or associated with a small cluster.  It may be possible to
substantiate this with deep Keck/LRIS imaging, which may be more
sensitive than HST to low surface brightness features.  Thirdly, the
existence of other galaxies in the HDF similar to c.~0 allows us to
identify other potential lenses and to repeat this procedure.  By far
the best candidate after \HDF\ is
\HDFx, shown in Figure~4.  If we are able to identify
several more instances of strong galaxy-galaxy lensing in this or
other fields, we will be able to derive better estimates of evolving
galaxy luminosity functions. This would be a large step toward solving
the puzzle of the excess faint sources.

\acknowledgements
We thank the Hubble Deep Field team, led by Bob Williams, for
planning, taking, reducing, and making public the phenomenal images of
the HDF.  We benefited from helpful conversations with Lee Armus,
Andrew Baker, Judy Cohen, Mark Dickinson, Mauro Giavalisco, Richard
Hook, Jordi Miralda-Escud\'e, Gerry Neugebauer and Chuck Steidel.  Lee
Armus, Len Cowie, Keith Matthews and Tom Soifer generously provided us
with near-infrared data in advance of publication.  We are grateful
for financial support from the U.S. National Science Foundation.  This
research made use of NASA's Astrophysics Data System Abstract Service.

\begin{figure}
\plotone{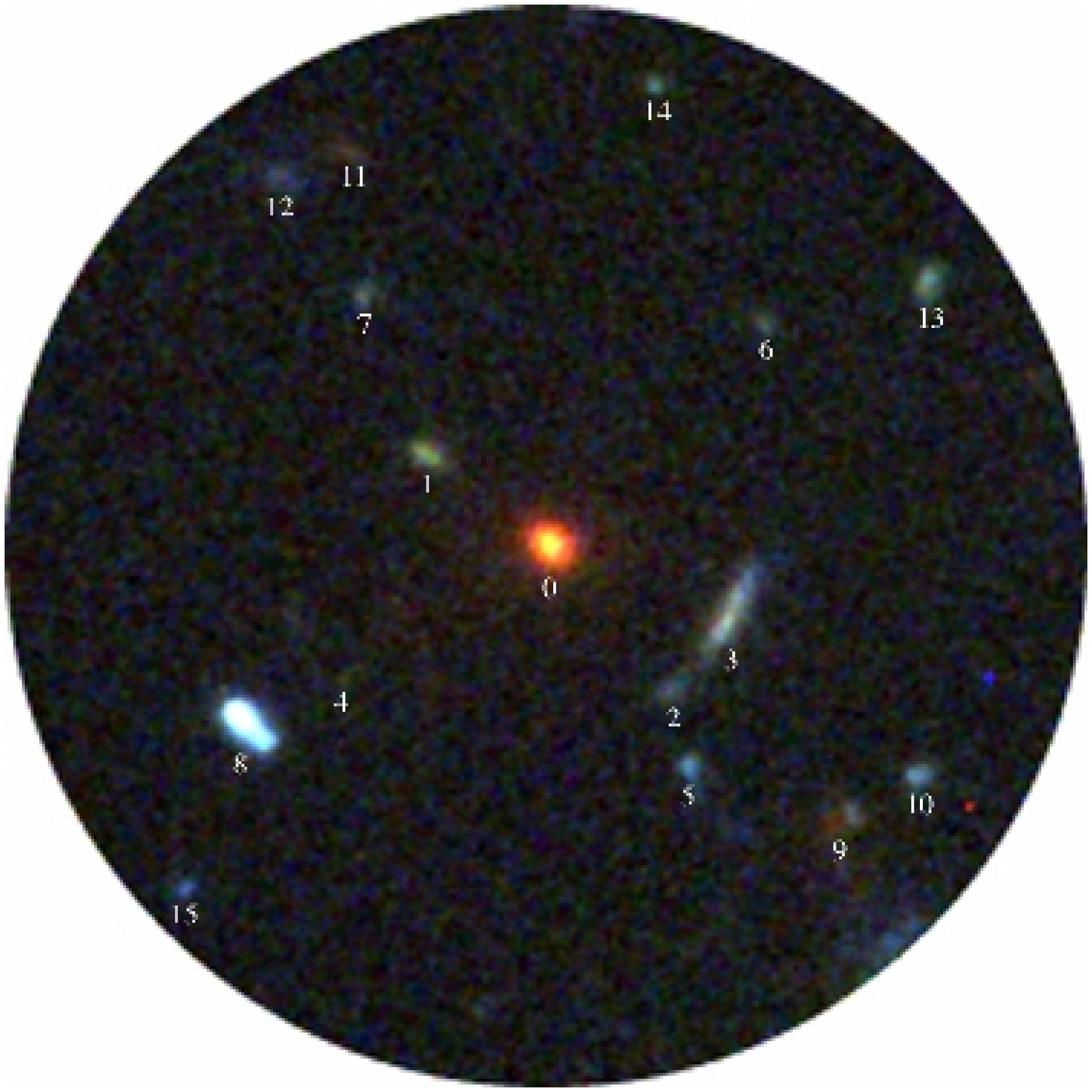}
\caption{
The 5~arcsec radius region centered on c.~0 of \HDF, with north up.
This is a true-color representation of the F450W, F606W and F814W
images.  The 16 components, numbered in order of radial distance from
c.~0, are marked.}
\end{figure}

\begin{figure}
\plotone{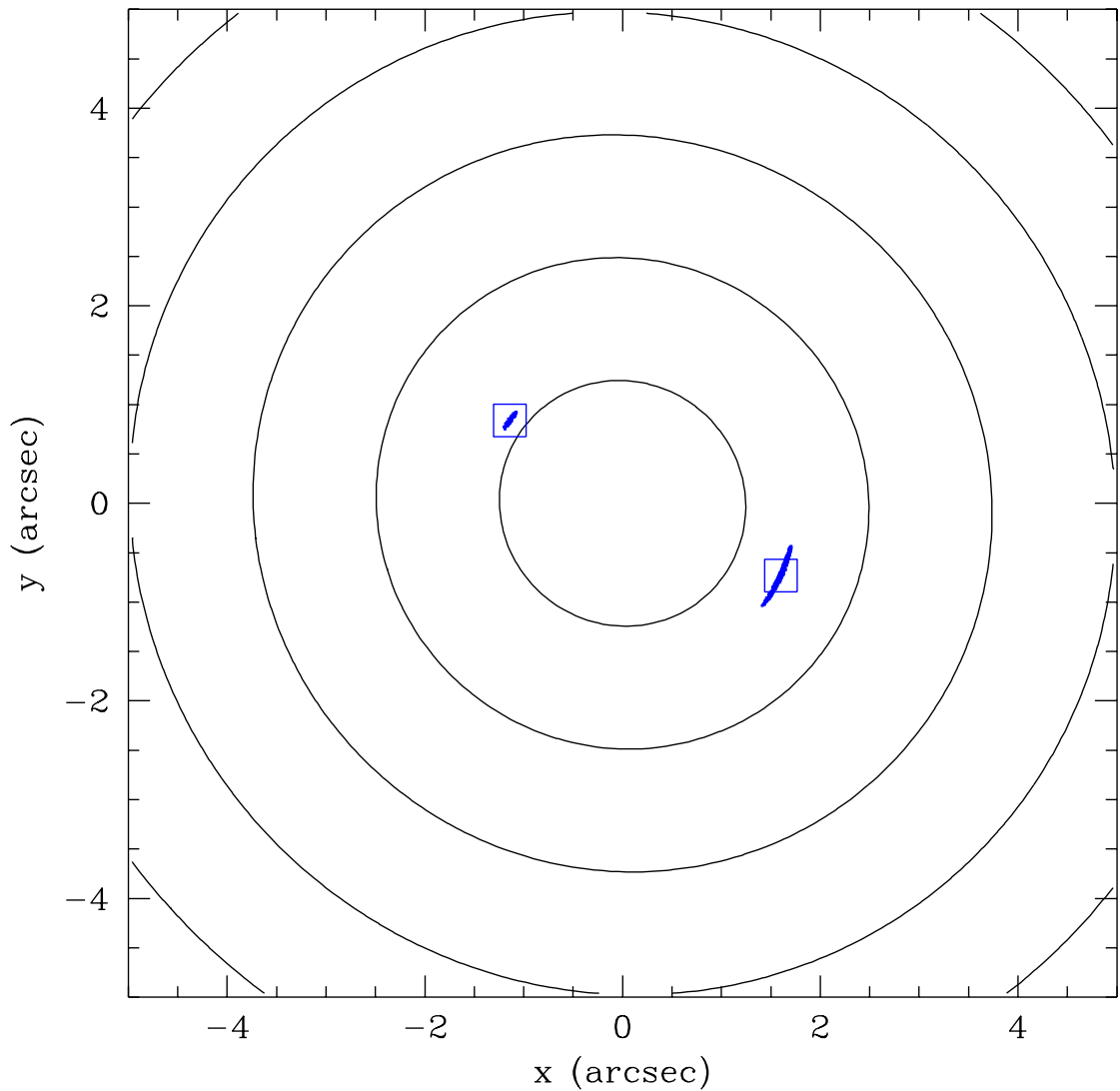}
\caption{
A model of cc.~1 and 3 of \HDF.  Observed positions are shown as boxes
and the elliptical blobs are images of circular source A.  The
parameters are $b=1.61$~arcsec, $\eta=0.021$~arcsec,
$\theta_{\eta}=47$~deg (N through E), and $r_A= 0.02$~arcsec.}
%besta=        1.607095
%bestgcel=     -0.001532
%bestgsel=     0.021253
%(bestgel=     0.021309)
%(bestthetael= -42.938489)
%bests=        0.000000
%bestchisq=    0.000001
%bestbeta[0]=  1.000000
\end{figure}

\begin{figure}
\plotone{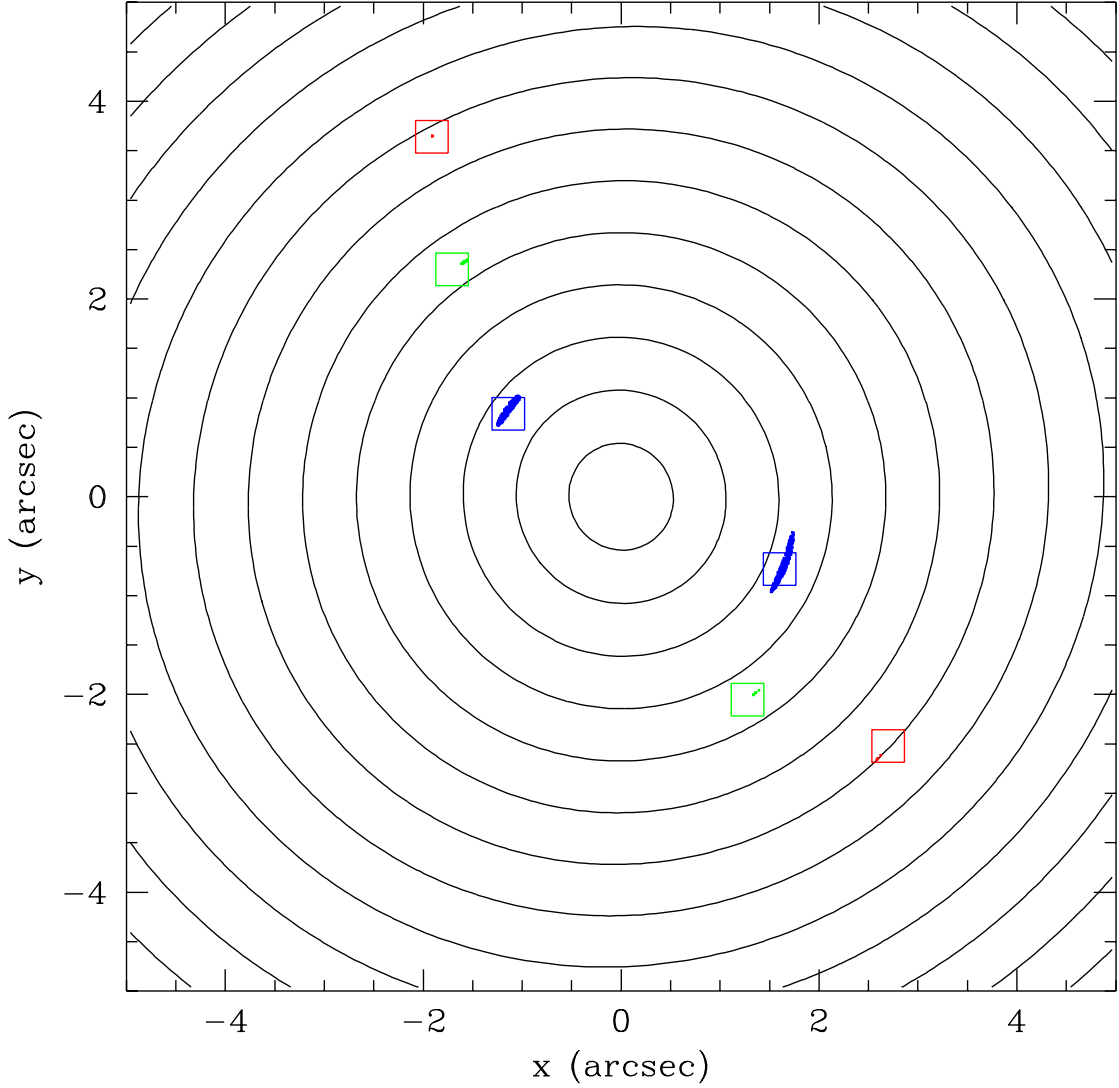}
\caption{
A model of cc.~1, 3, 9, 11, 5 and 7 of \HDF.  Observed positions are
shown as boxes and the elliptical blobs are images of circular sources
A (blue), B (red), and C (green).  The parameters are $b=3.74$~arcsec,
$\eta=0.090$~arcsec, $\theta_{\eta}=27$~deg (N through E),
$\gamma=0.070$, $\theta_{\gamma}=156$~deg, $\beta_B=2.31\,\beta_A$,
$\beta_C=1.58\,\beta_A$, $r_A=0.03$~arcsec, $r_B=0.0015$~arcsec and
$r_C=0.005$~arcsec.  Because cc.~9 and 7 are unresolved in the HDF
image, $r_B$ and $r_C$ are upper limits.}
%besta=        3.740741
%bestgcel=     0.053891
%bestgsel=     0.071886
%(bestgel=     0.089844)
%(bestthetael= -63.428957)
%bestgcex=     -0.041971
%bestgsex=     -0.056080
%(bestgex=     0.070046)
%(bestthetaex= 26.594010)
%bestchisq=    66.494920
%bestbeta[0]=  1.000000
%bestbeta[1]=  0.430832
%bestbeta[2]=  0.686267
\end{figure}

\begin{figure}
\plotone{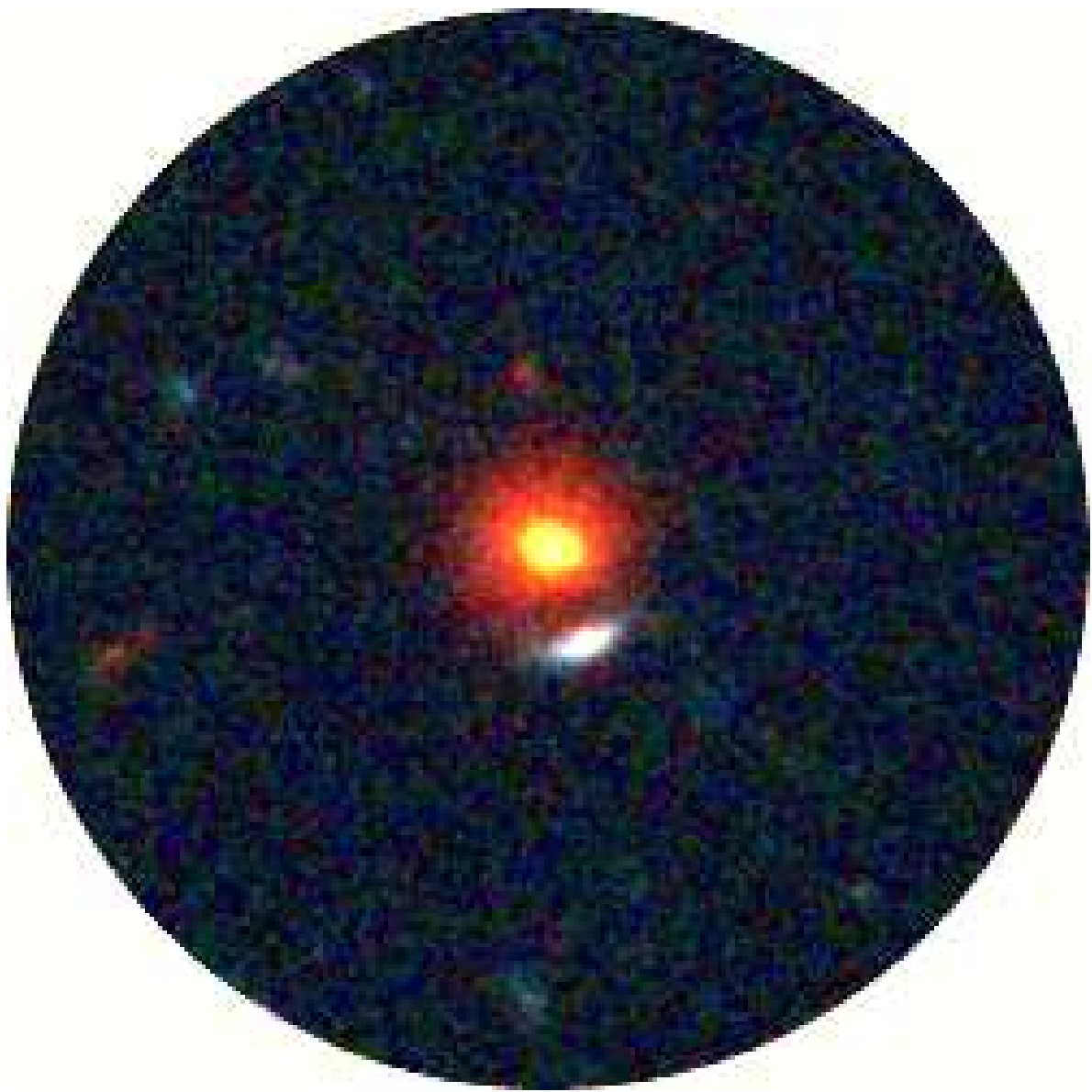}
\caption{
The 5~arcsec radius region centered on the central component
of \HDFx, the second-best lens candidate found so far in the HDF, with
north up.  This is a true-color representation of the F450W, F606W and
F814W images, stretched the same as Figure~1.}
\end{figure}

\end{document}